\documentclass[a4paper,11pt]{article}

\usepackage{pos}

\usepackage{framed}
\usepackage{graphicx}
\usepackage{color}
\usepackage[most]{tcolorbox}
\usepackage{bbold}
\usepackage{bbm}
\usepackage[normalem]{ulem}
\usepackage{rotating} 
\usepackage{slashed} 
\usepackage{cancel} 
\usepackage{braket} 
\usepackage{amstext}
\usepackage{array}
\usepackage{overpic}
\usepackage{bbm}
\usepackage{lipsum}

\newcommand{\seq}[1]{\begin{align*}#1\end{align*}}

\newcommand{\cref}[1]{\S~\ref{#1}}

\usepackage{dsfont}

\newcommand{\ignore}[1]{}

\title{A Fuzzy Axiverse from String Theory}

\author*[a,b]{Andreas Schachner}
		
\affiliation[a]{Arnold Sommerfeld Center for Theoretical Physics, Ludwig-Maximilian-University Munich,\\
  Theresienstr. 37, 80333 Munich, Germany}

\affiliation[b]{Department of Physics, Cornell University, Ithaca, NY 14853 USA}

\emailAdd{a.schachner@lmu.de}
\emailAdd{as3475@cornell.edu}

\abstract{
In this talk, I showcase models for fuzzy axion dark matter within the framework of type IIB string theory, focusing on axions originating from the Ramond-Ramond four-form in compactifications on Calabi-Yau orientifold hypersurfaces. These models are amenable to cosmological tests if a substantial relic abundance of fuzzy dark matter is produced.
I present a topologically exhaustive ensemble of more than 350{,}000 Calabi-Yau compactifications with up to seven axions together with a systematic analysis of the misalignment production of fuzzy dark matter.
The resulting dark matter composition is generally a mixture of fuzzy axions and heavier axions, including the QCD axion.
Dark photons frequently emerge due to the orientifold projection.
I will also comment on applications of optimisation strategies based on automatic differentiation for exploring the string axiverse.
This talk is partially based on \cite{Sheridan:2024vtt}.

}

\FullConference{2nd Training School and General Meeting of the COST Action COSMIC WISPers  (CA21106) (COSMICWISPers2024)\\
 10-14 June 2024 and 3-6 September 2024\\
Ljubljana (Slovenia) and Istanbul (Turkey)\\}

\begin{document}
\maketitle

\section{Introduction}
\label{sec:intro}

The string axiverse emerges naturally from string theory predicting a plenitude of axion-like particles~\cite{Svrcek:2006yi,Arvanitaki:2009fg,Cicoli:2012sz} arising from the compactification of extra dimensions. These axions span a wide range of masses and couplings, offering a rich phenomenological playground that connects high-energy string theory to low-energy cosmology and particle physics. Among the most compelling applications is \emph{fuzzy dark matter} (DM), a paradigm where an ultralight axion with a mass $m$ in the range $10^{-33}\text{ eV} \lesssim m \lesssim 10^{-18}\,\mathrm{eV}$ serves as a viable dark matter candidate.

\section{Fuzzy dark matter in Type IIB compactifications}\label{sec:ModelConst}

In this work we investigate fuzzy DM from the Ramond-Ramond (RR) 4-form in type IIB superstring theory.
The setting for our work is the \emph{Kreuzer-Skarke (KS) Axiverse}~\cite{Demirtas:2018akl}: axion theories arising in compactifications of type IIB string theory on Calabi-Yau (CY) orientifold hypersurfaces in toric varieties obtained from triangulations of four-dimensional reflexive polytopes \cite{Kreuzer:2000xy}.
We characterise the fuzzy axiverse for all orientifolds induced by suitable involutions of the ambient variety \cite{Moritz:2023jdb}, for all triangulations of fav. polytopes in the KS list with $2\le h^{1,1} \le 7$.
We will focus on orientifolds with $h^{1,1}_{-}=0$ for which $h^{1,1}=h^{1,1}_{+}$ counts the number of RR 4-form axions $\phi_i$.

In this setting,
the continuous shift symmetry of the axions $\phi_i$ is broken to a discrete one through Euclidean D3-instantons wrapping divisors $D_{A}=Q_A^i \, \tau_i$ with $\tau_{i}=\mathrm{Vol}(D_i)$ the divisor volumes of a basis of prime toric divisor $\lbrace D_{i}\rbrace$, $i=1,\ldots,h^{1,1}$, and $Q_{A}^i$ the instanton charge matrix.
The induced potential for the moduli $\tau_{i}$ and axions $\phi_{i}$ is given by\footnote{We choose for the flux superpotential $W_0 = 1$, for the string coupling $g_{s}=0.5$ and for the Pfaffians $A_{i}=1$.}
\vspace*{-0.2cm}
\begin{equation}\label{eq:potential} 
V(\tau_{i},\phi_{i}) =V_{\text{moduli}}(\tau_{i})+ \sum_{A}\Lambda^4_{A}\left(1-\cos(2\pi Q_{A}^i\phi_i)\right)+\ldots\;\, ,\quad \Lambda_{A}^4 \sim \, e^{K/2}|W| \frac{Q_A^i \, \tau_i}{\mathcal{V}} e^{-2\pi Q_A^i \, \tau_i} 
\vspace*{-0.3cm}
\end{equation}
with $\Lambda_{A}^4$ the instanton scales.
Throughout, we assume that moduli can be stabilised everywhere inside the Kähler cone $K_{X}$ via $V_{\text{moduli}}(\tau_{i})$, with masses larger than those of all axions $\phi_{i}$.
Importantly, we restrict our analysis to the geometric regime $\tau_{i}\geq 1$ in which the $\alpha'$ expansion is well-controlled.
Further, 
we compute axion masses $m_{i}$ and decay constants $f_{i}$ from \eqref{eq:potential}
in a hierarchical approximation where $\Lambda_{A}^{4}\gg\Lambda_{A+1}^{4}$ in a suitable ordering.
We then call an axion of mass $10^{-33}\text{ eV} \lesssim m_i \lesssim 10^{-18}\,\mathrm{eV}$ \emph{fuzzy} if it makes a significant contribution to the total DM abundance $\Omega_{\mathrm{DM}}$, see \cite{Sheridan:2024vtt} for detailed expressions for the relic abundances $\Omega_i$.
Following \cite{Gendler:2023kjt}, we also assume that the Standard Model can be realised on D7-branes hosted on suitable intersecting divisors. The QCD coupling $\alpha_{QCD}$ is set in the UV by $\tau_{QCD}=\text{Vol}(D_{QCD})$ where $D_{QCD}$ is the divisor hosting the QCD D7-stack.

With these assumptions in mind, we search our list of CY orientifold hypersurfaces for compactifications in which the fuzzy DM abundance is near the observational limits.
Specifically, we compute the DM abundance of all axions in the spectrum from vacuum realignment.
The resulting models contain one or multiple fuzzy axions, and a QCD axion.
Avoiding DM overabundance then typically requires changing the reheating temperature $T_{R}$, or tuning initial misalignment angles $\theta_{i}$. 
We note that fuzzy DM in Type IIB compactifications was previously investigated in~\cite{Cicoli:2021gss} by examining the fuzzy abundance in two canonical moduli stabilisation procedures.
This work complements \cite{Cicoli:2021gss} by working in explicit CY orientifolds and imposing the existence of a QCD axion, while not explicitly incorporating moduli stabilisation.

In what follows, we present two examples of axion EFTs arising in Type IIB compactifications containing both a fuzzy and a QCD axion.
The data needed to reproduce these examples can be found in a dedicated \href{https://github.com/sheride/fuzzy_axions}{GitHub repository}, see also \cite{Sheridan:2024vtt} for additional examples and their cosmologies.

\subsection{Illustrative example --- $h^{1,1}=2$}\label{ex:h11_2}

\begin{figure}[t]
    \centering 
    \includegraphics[width=0.75\linewidth , height=5.5cm]{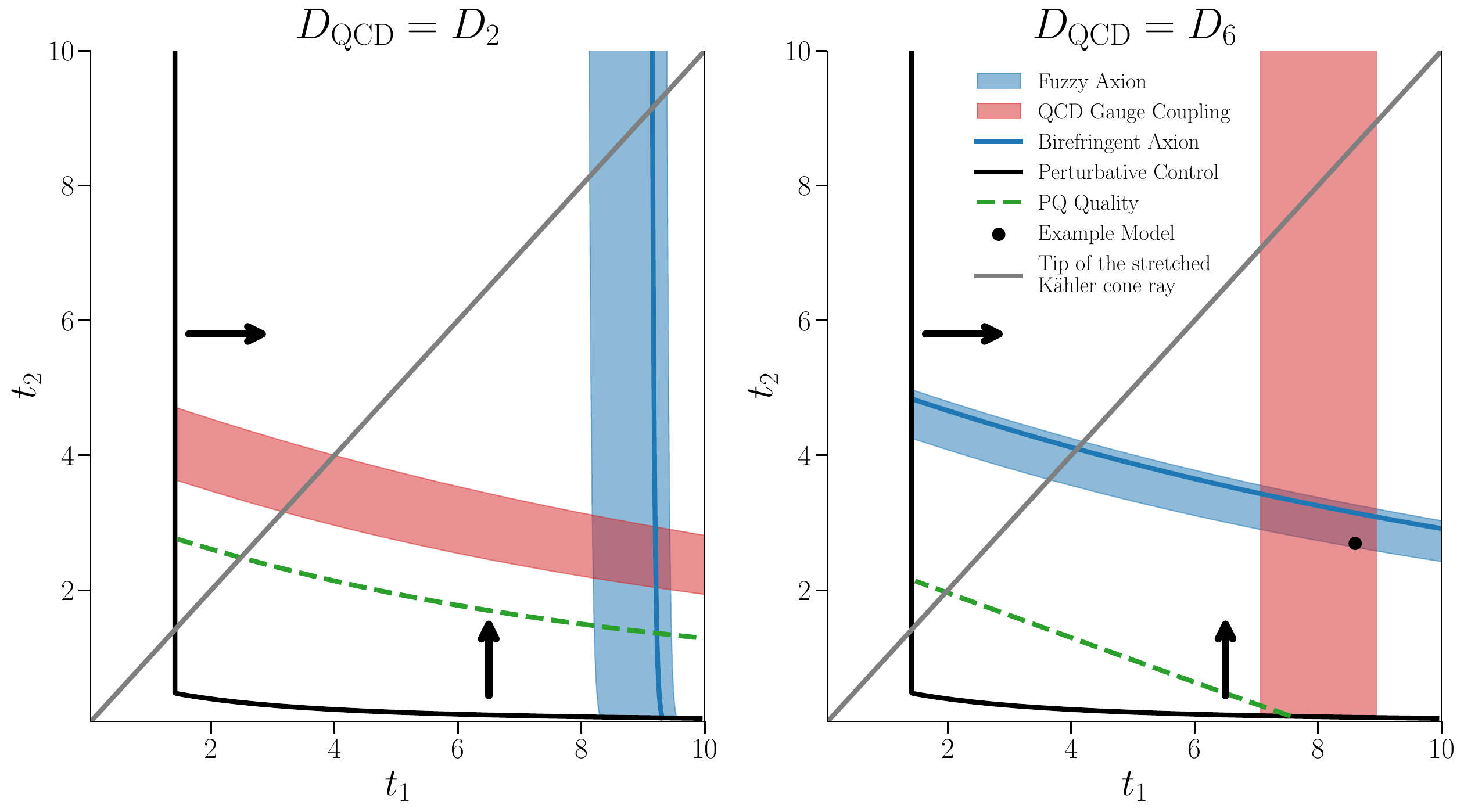} 
    \caption{
    K\"ahler cone for the geometry with $h^{1,1} = 2$ for two choices of divisors hosting QCD. 
    }
    \label{fig:h_2_cone_v2}
\end{figure}

We begin by illustrating how fuzzy and QCD axions arise in particular regions of $K_{X}$.
We consider a CY orientifold with $(h^{1,1},h^{2,1})=(h^{1,1}_+, h^{2,1}_-) = (2, 132) $ denoting the prime toric divisors $D_1, \ldots, D_{6}$.
First,
we explore the full K\"ahler moduli space in Fig.~\ref{fig:h_2_cone_v2}. 
We host QCD on the $h^{1,1} = 2$ most relevant instantons associated to $D_2$, $D_6$.
In red we identify regions where $\tau_{QCD}$ is consistent with the experimentally observed QCD gauge coupling $\alpha_\mathrm{QCD}(m_Z)$. 
In blue, we delineate the region where an axion has mass $m$ satisfying  $10^{-33} \; \mathrm{eV} \leq m \leq 10^{-18} \; \mathrm{eV}$.
The gray curve in Fig.~\ref{fig:h_2_cone_v2} denotes the ray generated by the tip of the stretched K\"ahler cone
which does not intersect the region in moduli space where a fuzzy axion and a QCD axion can be simultaneously realised.
This highlights the sacrifice one makes by restricting to a dimension-one submanifold of moduli space.

Let us provide an explicit example by fixing K\"ahler parameters $\mathbf{t}_\star$ corresponding to the black dot in the right panel of Fig.~\ref{fig:h_2_cone_v2}.
Assuming moduli stabilisation at $\mathbf{{t}}_\star$, 
we find
\vspace*{-0.2cm}
\begin{equation}
\tau_\mathrm{QCD} = 37\, ,\quad m_i = (1.8 \cdot 10^{-9}, 5.0 \cdot 10^{-20})\; \mathrm{eV}\, ,\quad f_i = (3.2 \cdot 10^{15}, 1.2 \cdot 10^{16})\; \mathrm{GeV}\,.
\vspace*{-0.2cm}
\end{equation}
In particular, the first axion is the QCD axion associated to $D_6$, and the second axion is the fuzzy axion associated to $D_2$. The choice of initial misalignment angles $\theta_{i}$ that describes the observed dark matter abundance is $\theta_{i} = (0.0063, 1)$ leading to $\Omega_i/\Omega_\mathrm{DM} = (0.5, 0.5)$. 
Cosmologically, because we have no other (heavy) axions, we have no further constraints on e.g. $T_R$.

\subsection{Lightest fuzzy abundance --- $h^{1,1}=7$}\label{sec:best_abundance}
 
\begin{figure}
    \centering
    \includegraphics[width=0.7\linewidth]{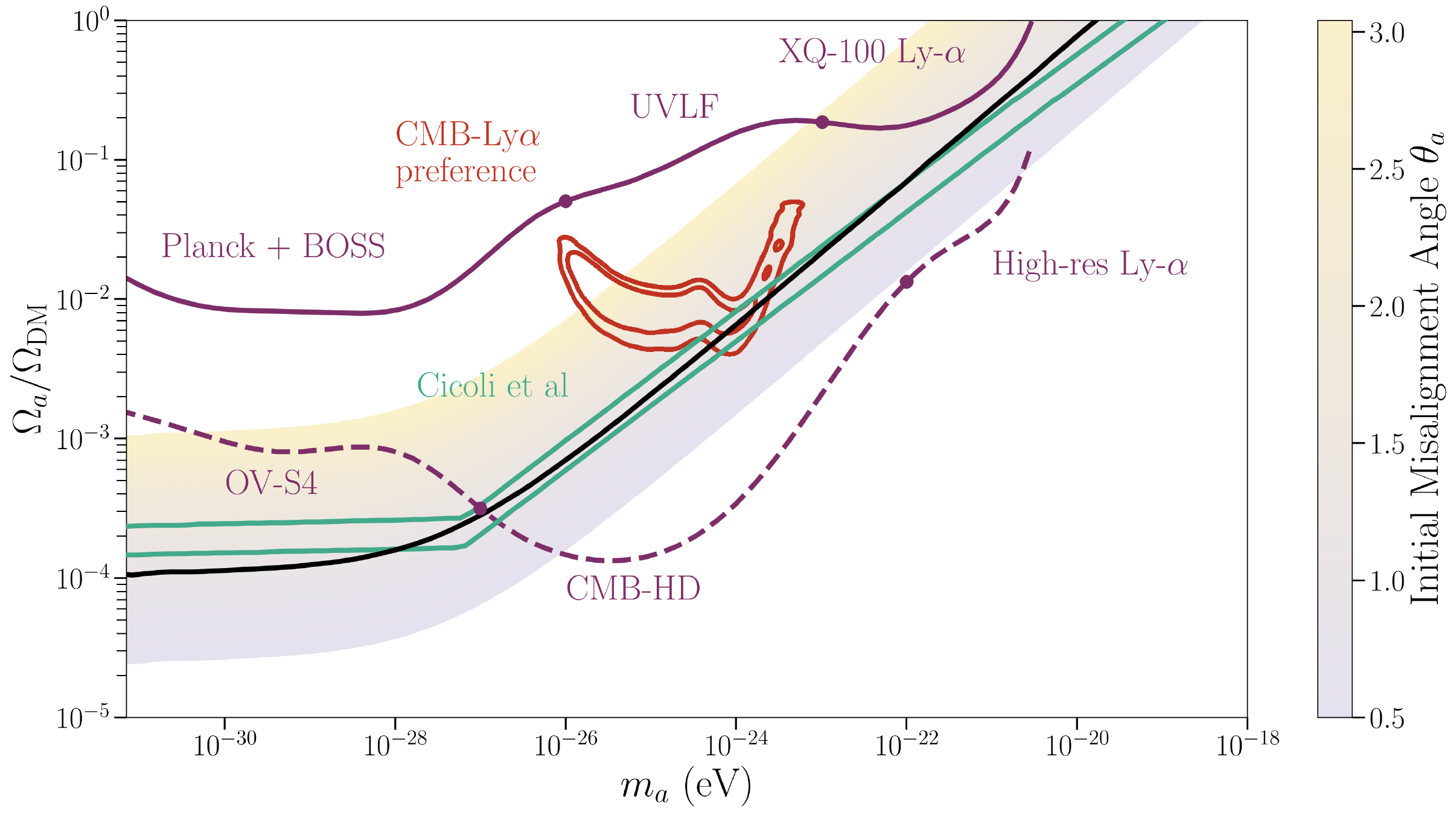}    
    \caption{Relic abundance as a function of the axion mass for different initial misalignment angles, overlaid by the abundance prediction from \cite{Cicoli:2021gss}.
     }
    \label{fig:max_abundance}
\end{figure}

Next, we present the geometry containing the lightest fuzzy axion found in our setting constituting 100\% of the observed DM.
The corresponding CY orientifold has $(h^{1,1}, h^{2,1}_{+}, h^{2,1}_{-}) = (7, 9, 16)$ with
the QCD and fuzzy axion being associated to $D_1$ and $D_2$ respectively.
In this example, the K\"ahler parameters $\mathbf{t}_\star$ are obtained by exploring $K_{X}$ for points with optimal fuzzy misalignment abundance. 
Indeed, computing the axion DM abundance involves solving cosmological evolution equations that depend on initial conditions and Kähler parameters.
By implementing differentiable code using the \textsc{jax}~library~\cite{jax2018github},
we can use
auto-differentiation (AD) to efficiently compute gradients of a carefully designed loss function that encodes the desired geometric properties.
This enables the use of gradient-based optimisation algorithms to 
efficiently identify optimal K\"ahler parameters where the untuned misalignment abundance matches the observed value.

In this way, we find K\"ahler parameters $\mathbf{t}_\star$ as specified in the supplementary materials \cite{Sheridan:2024vtt}.
Here, we compute the KK scale $m_{KK} = 6.9 \cdot 10^{16}$ GeV, the QCD divisor volume $\tau_\mathrm{QCD} = 35.6$, and the axion masses $m_i$ and decay constants $f_i$ for the two lightest axions
\vspace*{-0.2cm}
\begin{equation}
\begin{aligned}
m_i &= (\ldots,7.7 \cdot 10^{-10}, 1.7 \cdot 10^{-20})\; \mathrm{eV}\,, \quad f_i = (\ldots,7.4 \cdot 10^{15}, 2.2 \cdot 10^{16})\; \mathrm{GeV}\,. 
\end{aligned}
\vspace*{-0.2cm}
\end{equation} 
In particular, the heavy axion is the QCD axion and the light axion is the fuzzy axion. Choosing $\theta_{i}=1$ for the fuzzy axion, its abundance satisfies $\Omega_i/\Omega_\mathrm{DM} = 1$. This model yields the lightest axion whose untuned misalignment abundance achieves the observed value, though heavier axions in this model will in general overproduce DM.

This maximal fuzzy axion misalignment abundance is compared against theoretical predictions along with phenomenological constraints (solid), forecasts (dashed), and targets in Fig.~\ref{fig:max_abundance}, see \cite{Sheridan:2024vtt} for a detailed discussion. The shaded region is produced by dilating the K\"ahler parameters along the ray generated by $\mathbf{t}_\star$ 
and the initial misalignment angle.
We observe reasonable agreement between our explicit model and the predictions for four-form axions given in \cite{Cicoli:2021gss}, with discrepancies between the slopes being attributed to the different samplings of loci in moduli space. 
Interestingly, the predictions of \cite{Cicoli:2021gss} obtained by imposing moduli stabilisation largely overlap with the black line for the untuned misalignment abundance in Fig.~\ref{fig:max_abundance}.

\section{Conclusions and future directions}

We have examined the prevalence of fuzzy axion DM in Type IIB compactifications on CY orientifold hypersurfaces. Our main finding is that,in certain regions of the landscape, an observably-large fuzzy DM relic density is indeed feasible within this framework.
However, this is usually accompanied by an overabundance of heavier axion DM. 
This is remedied by either finding configurations with hierarchical decay constants, with $f_{\text{other}} \ll f_{\text{fuzzy}}$,
or, for example, by reducing the initial misalignment angles $\theta_{i}$ and/or the reheating temperature $T_{R}$ (see \cite{Sheridan:2024vtt} for details). Moreover, our analysis of the statistics of orientifold projections indicates that dark photon fields are generically present in the KS axiverse, and they merit further scrutiny.

In the future, recent applications of \emph{Genetic Algorithms}, as explored in \cite{MacFadden:2024him}, present a promising direction for studying the KS axiverse. This method excels by optimising a broad class of EFTs derived from Type IIB CY compactifications, evolving them to e.g. enhance axion-photon couplings.
In the same vein, automatic differentiation (AD) is an invaluable tool for addressing the computationally intensive optimisation problems in the string axiverse.
By enabling precise and efficient derivative calculations for observables such as DM abundance or moduli potentials (see in particular \cite{Dubey:2023dvu}),
AD facilitates a systematic exploration of the parameter space of string compactifications when combined with suitable optimisation strategies.
In this way, we can begin to connect low-energy physics to high-energy theory by, for instance, identifying axions that align with cosmological and astrophysical observations.

\textbf{Acknowledgements.} I would like to thank Federico Carta, Naomi Gendler, Mudit Jain, David J.~E.~Marsh, Liam McAllister, Nicole Righi, Keir K. Rogers, and Elijah Sheridan for collaboration on this project.
This article is based upon work from COST Action COSMIC WISPers CA21106,
supported by COST (European Cooperation in Science and Technology).

\bibliographystyle{utphys}
\bibliography{Literatur}

\end{document}